\documentclass[10pt,conference,comsoc]{IEEEtran}
\IEEEoverridecommandlockouts

\usepackage{cite}
\usepackage{amsmath,amssymb,amsfonts}
\usepackage{algorithmic}
\usepackage{algorithm}
\usepackage{float}
\usepackage{graphicx}
\usepackage{textcomp}
\usepackage{xcolor}
\usepackage{tcolorbox}
\tcbuselibrary{skins, breakable}
\usepackage{balance}
\usepackage{booktabs}
\usepackage{multirow}
\usepackage{url}
\usepackage{tikz}
\usepackage[table]{xcolor}
\usepackage{tabularx}
\usepackage{comment}
\usepackage{enumitem}
\usepackage{pgfplots}
\pgfplotsset{compat=1.17}
\usepackage{listings}
\definecolor{TableHeader}{HTML}{2F5597}
\definecolor{TableStripe}{HTML}{EAF1F8}
\definecolor{TableHighlight}{HTML}{FFF2CC}
\tcbset{
  colback=gray!5,
  colframe=black!50,
  arc=2mm,
  boxrule=0.5pt,
  left=3pt,right=3pt,top=3pt,bottom=3pt
}

\usetikzlibrary{arrows.meta,backgrounds,fit}
\usetikzlibrary{shapes.geometric, arrows.meta, positioning, backgrounds}
\def\BibTeX{{\rm B\kern-.05em{\sc i\kern-.025em b}\kern-.08em
    T\kern-.1667em\lower.7ex\hbox{E}\kern-.125emX}}

\usepackage{listings}
\begin{document}
\title{ NDT Factory: Synthesizing Verified Network Digital Twins from Semantic Models via Multi-Agent LLM\vspace{-0.1in}}
%\title{On-Demand Behavioral Digital Twin Generation for Autonomous Networks via Semantic Model-Driven Synthesis}

\author{
\IEEEauthorblockN{Sudipta Acharya$^1$, Petar Djukic$^2$, Burak Kantarci$^1$\vspace{-0.1in}}\\
\IEEEauthorblockA{\textit{$^1$University of Ottawa, Ottawa, ON, Canada}\\
\textit{$^2$Bell Labs Research, 600 March Road,
Kanata, ON K2K 2E6, Canada}\\
$^1$\{sacharya2, burak.kantarci\}@uottawa.ca,~$^2$petar.djukic@nokia-bell-labs.com}
\vspace{-0.4in}}

\maketitle
\thispagestyle{empty}
\pagestyle{empty}

% ---------------------------------------------------------------
\begin{abstract}
Autonomous network management requires systems that can evaluate Network Service Intents (NSIs) under varying conditions without manual implementation of analysis logic, as envisioned in TM Forum Level~4 (L4) autonomy. Behavioral Network Digital Twins (NDTs) enable such evaluation, but existing NDTs rely on pre-defined analytical logic, limiting adaptability for evolving closed-loop control. This paper introduces the \emph{NDT factory}, a multi-agent software system that synthesizes executable behavioral NDTs on demand from semantic models using Large Language Model (LLM). We validate the system using a Call Admission Control (CAC) case study, where deterministic \emph{what-if} analysis serves as the admission decision process. The NDT factory generates a complete CAC NDT through parallel synthesis and orchestration, achieving 100\% compilation and test pass rates across multiple runs. Simulation over 300 NSIs shows 99.3\% decision agreement with a reference implementation, 90\% admission rate, and correct attribution of all rejections, demonstrating reliable synthesis with deterministic, verifiable execution.
\end{abstract}


% ---------------------------------------------------------------
\section{Introduction}
\label{sec:intro}

Network Service Intents (NSIs) are declarative, operator-issued specifications that express high-level service requirements without prescribing implementation details. Autonomous network management, as envisioned by TM Forum Level~4 (L4) autonomy~\cite{tmf_ig1251}, the
Experiential Networked Intelligence (ENI) framework of the European Telecommunications Standards Institute
(ETSI)~\cite{etsi_eni_051}, and Djukic et al.~\cite{11613137}, requires networks to process NSIs and make admission decisions autonomously, while operators retain a supervisory monitoring role. A key challenge is determining how service request admission will affect network state and whether it may violate latency, utilization, or existing-service constraints. This work focuses on behavioral Network Digital Twins (NDTs), which prospectively apply a service request to a virtual state, recompute affected metrics, and verify request and existing-service constraints before modifying the live network.

However, existing behavioral NDTs rely on predefined analytical logic fixed at design time as hand-coded algorithms or trained models~\cite{verdecchia2024network, li2025generative, 3gpp_tr28915}. This logic depends on the constraint semantics of an NSI. For example, bandwidth and latency requirements demand residual-capacity and path-delay analysis, whereas availability requirements may require redundant or disjoint-path evaluation. Requests sharing the same constraint-evaluation logic can reuse an NDT, but new constraint types, dependencies, or decision policies may require new evaluation modules. Consequently, evolving NSIs require repeated manual engineering, tightly coupling an intent's declared outcome with the logic used to evaluate it. This limits scalable closed-loop L4 autonomy~\cite{11613137} and motivates the automatic synthesis of behavioral NDTs from declarative specifications.

In this paper, we propose the \emph{NDT factory}, a multi-agent software system that synthesizes executable behavioral NDTs on demand. When a required NDT is absent, it generates behavioral evaluation logic through Large Language Model (LLM)-based code synthesis, validates it using a \emph{compile--test--debug} loop, and caches the deterministic executable for reuse. Its central abstraction is the \emph{semantic model}, a structured YAML specification of the inputs, operations, constraints, and validation cases of the target behavior without prescribing its source-code implementation. At runtime, the cached NDT evaluates incoming NSIs against the current network state without LLM involvement. We validate the system using a Call Admission Control (CAC) case study, where a complete multi-component CAC NDT is synthesized from semantic models. The key contributions are:

\begin{enumerate}[leftmargin=*, itemsep=1pt, topsep=2pt, parsep=0pt, partopsep=0pt]
\item \textbf{\emph{Semantic model:}} A formal interface between intended NDT behavior and automatically generated management-plane executable logic.
\item \textbf{\emph{NDT Factory:}} A multi-agent system that translates semantic models into deterministic, compile-and-test-validated executables through coordinated LLM-based synthesis.
\item End-to-end synthesis of the CAC NDT from semantic models, implementing deterministic \emph{what-if} analysis, validated through discrete-event simulation on a realistic topology.
\end{enumerate}

The remainder is organized as follows: Section \ref{sec:related} reviews related work; Section \ref{sec:architecture} presents the NDT factory architecture and introduces the semantic model; Sections \ref{sec:casestudy}–\ref{sec:evaluation} cover the case study and evaluation; Section \ref{conclusion} concludes.
\begin{figure*}[t]
\centering
\includegraphics[width=0.61\textwidth]{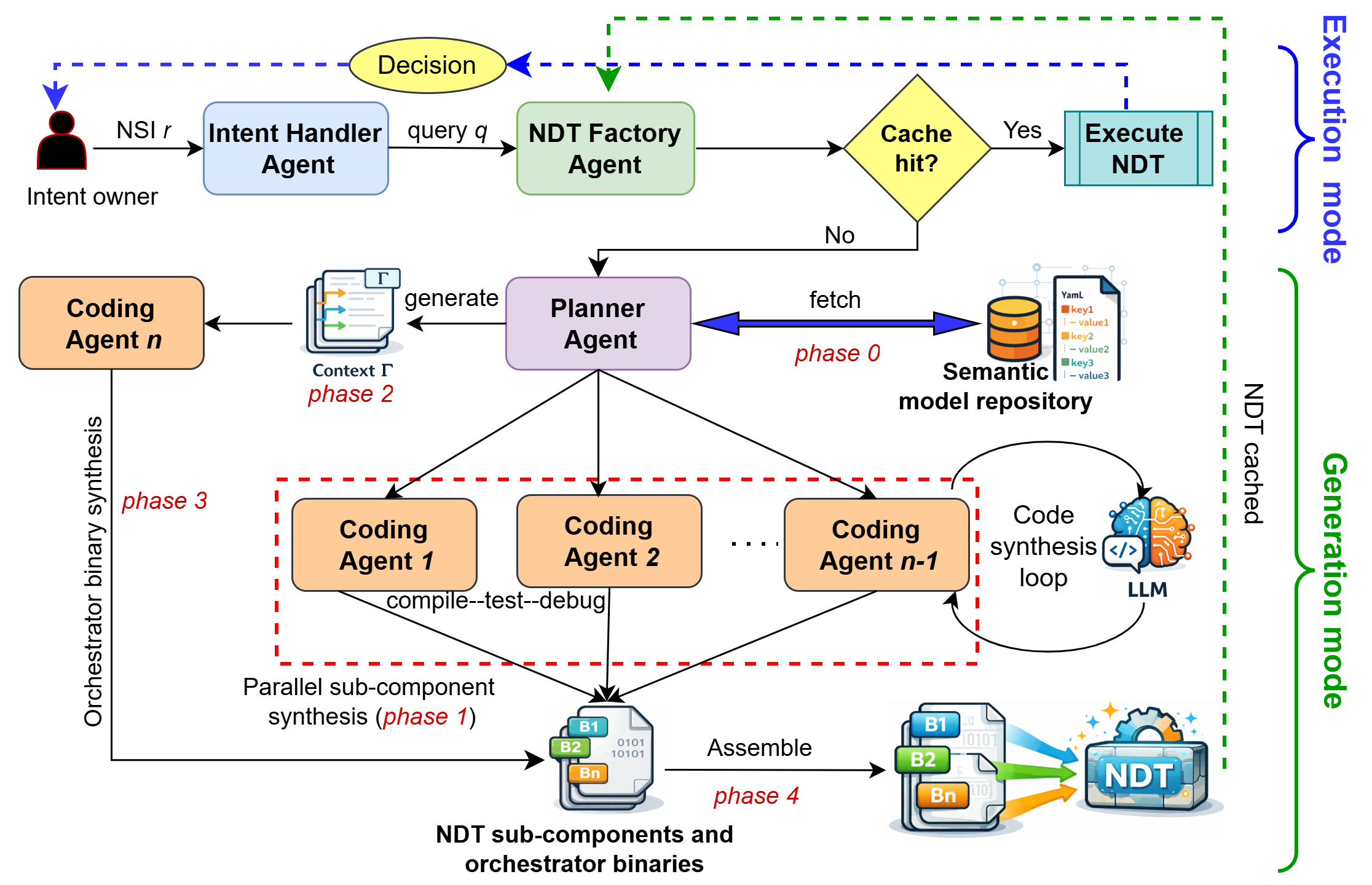}
\caption{Architecture of the NDT factory: in execution mode, a cached NDT produces decisions; in generation mode, the Planner coordinates parallel Coding agents to synthesize sub-components from semantic models, assemble the pipeline, and cache it for reuse \vspace{-0.2in}}
\label{fig:architecture}
\end{figure*}

%\noteBK{what we've found out of these contributions? also a brief explanation on the paper structure needed}
%
\section{Related Work and Motivation}
\label{sec:related}
Recent surveys establish NDTs as core infrastructure for network planning and control in 5G/6G systems~\cite{verdecchia2024network,yang2024architectural}. Existing NDT studies focus on architecture, analytics, or scenario-specific simulation rather than automatic behavioral synthesis. Ak \emph{et al.} develop \emph{what-if} frameworks using data-driven evaluation~\cite{ak2024if} that rely on manually designed logic tied to specific scenarios. Recent 6G NDT architectures similarly assume behavioral modules are engineered separately~\cite{yang2024architectural}. Generative Artificial Intelligence (AI) approaches enhance data-driven analysis and decision support within NDTs~\cite{li2025generative,muhammad2024integrating} while on-demand synthesis of executable logic for unseen intent types remains an open research opportunity. Recent work by Djukic et al.~\cite{11613137} positions the digital twin factory as a management-plane agent that uses Large AI Models to interpret semantic models and synthesize NDTs in response to evolving network conditions. Building on this standards-oriented vision, we implement and evaluate an executable multi-agent pipeline for behavioral NDT synthesis.

Another research thread focuses on LLM-based code generation, which offers synthesis capabilities through multi-turn and collaborative frameworks~\cite{qian2024chatdev,hong2023metagpt,jiang2026survey}, with domain-specific extensions to networking~\cite{kan2024mobile}. While prior approaches treat code generation as a prompt-driven software development task, our objective is the synthesis of executable behavioral NDTs from semantic models that formally define domain inputs, operations, constraints, output conventions, and validation tests. In our system, the LLM is restricted to synthesis, while the compiled NDT executable performs runtime behavioral analysis. Prior NDT research focuses on architectures and scenario-specific \emph{what-if} analysis, while LLMs enable general synthesis. In contrast, our approach grounds synthesis in explicit behavioral semantics and makes the semantic model the primary driver, ensuring correctness through test-based validation.

\section{The proposed system: NDT factory}
\label{sec:architecture}

The NDT factory is a multi-agent software system composed of four interacting components: the \emph{Intent handler} agent, \emph{NDT factory} agent, \emph{Planner} agent, and a pool of parallel \emph{Coding} agents. The proposed system operates in two modes. In \emph{generation mode}, triggered when a required NDT is absent from the cache, the Planner and Coding agents collaboratively synthesize, compile, and validate components from semantic models to construct the target NDT. In \emph{execution mode}, for subsequent NSIs requiring the same analysis, the pre-generated NDT is invoked by the NDT factory agent, eliminating generation overhead. Once created, the NDT executes deterministically while preserving semantic constraints. Fig.~\ref{fig:architecture} illustrates the overall architecture and workflow of the NDT factory.

\paragraph{\textbf{Intent handler agent}}

The Intent handler is the entry point for incoming NSIs. It performs two functions. First, it parses and validates the NSI by enforcing syntactic constraints, including type, range, and consistency checks, rejecting malformed or incomplete requests. While this validation ensures that the intent is well-formed, it does not fully capture its semantic correctness or operational feasibility. Second, it transforms the validated intent into a structured behavioral analysis query (e.g., a \emph{what-if} evaluation query). This query is forwarded to the NDT factory agent.

\paragraph{\textbf{NDT factory agent}}
The NDT factory agent acts as the central coordinator of the software system. It receives the structured query from the Intent handler and manages execution of the behavioral NDT. Its responsibilities include:

\begin{enumerate}[leftmargin=*, itemsep=1pt, topsep=2pt, parsep=0pt, partopsep=0pt]
\item \textbf{NDT lookup}: Check whether the required NDT binary is available in the cache.
\item \textbf{On-demand generation}: If unavailable, invoke the Planner agent to synthesize, compile, and validate the NDT from the corresponding semantic model, and cache the resulting binary.
\item \textbf{Metadata collection}: Retrieve runtime network and service state, including topology (link capacities and utilization) and active services (paths, bandwidth allocations, and Service Level Agreement (SLA) constraints).
\item \textbf{NDT execution}: Construct the NDT input by combining the query with collected metadata, invoke the binary, and parse its structured output. To ensure consistency, it resolves the NDT input schema by aligning semantic model data structures with the generated code interfaces; this mapping is cached after the first invocation.
\item \textbf{State update and release}: Update the network state upon admission, and release allocated resources when services complete or expire.
\end{enumerate}

\paragraph{\textbf{Planner agent}}

Given a high-level orchestrator semantic model specifying the behavioral twin's components and dependencies, the Planner produces a compiled and verified executable NDT. Unlike conventional planners that decompose NSIs into data-plane network services, it operates in the management plane to synthesize an NDT service for behavioral intent evaluation. The system therefore distinguishes the data-plane network service from the management-plane NDT service. The Planner operates as follows:

\begin{description}
  \item[Phase 0: Parsing] Extract component (semantic models) dependencies from the orchestrator semantic model.

\item[Phase 1: Parallel generation] Retrieve the component semantic models, resolve dependencies, and distribute independent components among Coding agents for concurrent source-code generation and compilation.

\item[Phase 2: Context assembly] Build an orchestration context containing the generated components' binary interfaces, data-flow contracts, and pipeline order to support correct integration.

\item[Phase 3: Orchestrator generation] Generate the orchestrator source code via the Coding agent using the orchestrator semantic model and assembled context. The orchestrator acts as the main function of the NDT, coordinating all dependent components and producing the final decision.

\item[Phase 4: Assembly and verification] Assemble the components into a complete pipeline and verify its correctness and completeness, yielding the executable behavioral NDT.
\end{description}

\paragraph{\textbf{Coding agent}}
The Coding agent is the generative unit of the system. %Given a semantic model, it produces an executable implementation using a configurable LLM backend.
It performs the following tasks:

\begin{enumerate}[leftmargin=*, itemsep=1pt, topsep=2pt, parsep=0pt, partopsep=0pt]
\item Construct a structured prompt from the semantic model and optional context.
\item Invoke the LLM, which generates source code and executes tool calls (file writes, compilation, and test execution) within its agentic context. The LLM operates on both the prompt
and tool-call outputs.
\item Validate that the generated code compiles and passes all semantic-model-defined test cases.
\item On failure, incorporate compiler and test output as feedback into a new prompt and retry (up to four iterations).
\end{enumerate}

A synthesized code component is accepted only if it compiles successfully and passes all specified tests.

\paragraph*{\textbf{End-to-End Flow}}

The end-to-end workflow of the NDT factory is described below.
\begin{enumerate}[label=\textbf{Step \arabic*.}, leftmargin=*, align=left]

\item \textbf{Intent parsing:} The Intent handler receives an NSI, validates parameters, and transforms it into a structured behavioral query forwarded to the NDT factory.

\item \textbf{NDT lookup:} The NDT factory agent checks whether a compiled NDT for the required intent type exists in the cache.

\item \textbf{Execution mode:} If found, the NDT factory agent collects the current network and service states, invokes the cached NDT, and returns the admission decision without LLM involvement.

\item \textbf{Generation mode -- Sub-component synthesis:} Otherwise, the NDT factory agent invokes the Planner, which retrieves the orchestrator semantic model, extracts dependencies, and dispatches Coding agents in parallel to synthesize and validate sub-components via an LLM-driven compile--test--debug loop.
\item \textbf{Generation mode -- Context assembly:} The Planner constructs the orchestration context from generated sub-components and forwards it, along with the orchestrator semantic model, to a Coding agent.

\item \textbf{Generation mode -- Orchestrator synthesis:} The Coding agent synthesizes the orchestrator source code using the provided context and semantic model.

\item \textbf{Pipeline assembly and caching:} The Planner assembles and validates the sub-components and orchestrator, then returns the compiled NDT to the NDT factory agent, which caches the binary and proceeds as in Step~3.

\end{enumerate}

\begin{figure}[t]
\centering
\begin{tcolorbox}[
  colback=blue!3,
  colframe=blue!40,
  colbacktitle=blue!50,
  coltitle=white,
  title=\textbf{Semantic Model: Bandwidth Feasibility Pruner},
  fonttitle=\bfseries,
  boxsep=1pt,
  left=2pt,
  right=2pt,
  top=3pt,
  bottom=2pt,
  enhanced,
  sharp corners
]
\begin{lstlisting}[
  basicstyle=\ttfamily\scriptsize,
  lineskip=-1.2pt,
  aboveskip=2pt,
  belowskip=0pt
]
name: bandwidth_feasibility_pruner
description: Prune paths where bandwidth demand exceeds available link capacity

parameters:
  - name: topology
    type: NetworkTopology
  - name: request
    type: ServiceRequest
    fields: [src_node, dst_node, bandwidth_mbps]

operations:
  - name: PrunePaths
    description: Return paths with sufficient residual bandwidth on all links

conventions:
  exact_json_fields: [feasible_paths, pruned_count]

tests:
  - input: {bandwidth_mbps: 100}
    expected: {feasible_paths: [...]}
\end{lstlisting}
\end{tcolorbox}
\caption{Semantic model of the Bandwidth Feasibility Pruner (CAC case study)}
\label{fig:yaml}
\end{figure}
\paragraph*{\textbf{The Semantic Model}}
\label{sec:semantic}

The semantic model is a structured, verifiable YAML specification defining an NDT's intended behavior through its inputs, data structures, operations, constraints, output conventions, and test cases. It follows the parameter and data-structure conventions of the Internet Engineering Task Force (IETF) NDT architecture~\cite{zhou2024network} and the 3$^{rd}$ Generation Partnership Project (3GPP) TR~28.915~\cite{3gpp_tr28915}, facilitating integration with existing network models. For each target NDT, network-domain specialists define its logic, constraints, and outputs, while system engineers encode them as structured specifications and validation cases. Operationally, these models should undergo structural checks, expert review, and reference-case validation under version control, with each approved change triggering revalidation, cache invalidation, and NDT regeneration. Fig.~\ref{fig:yaml} presents an example of semantic model from the case study (Section~\ref{sec:casestudy}).

% ---------------------------------------------------------------

\section{Case Study}
\label{sec:casestudy}
We validate the system through a CAC case study in packet-switched networks. Given an incoming NSI, the CAC NDT applies a deterministic \emph{what-if} algorithm that projects the post-admission network state and evaluates bandwidth availability, link utilization, end-to-end latency, and existing-service SLAs. It admits the request only if all constraints hold; otherwise, it rejects it with per-constraint reasoning.

\subsection{Problem Formulation}

Let $G=(V,E)$ be a bidirected network graph, where each link $e\in E$ has capacity $C_e$ (Mbps), propagation delay $d_e$ (ms), and utilization $\rho_e\in[0,1)$. The active service set $\mathcal{S}=\{(P_k,B_k,L_k^{\max})\}$ records the path, reserved bandwidth, and latency SLA of each admitted service.
Let $\bar{s}=1$~kbit denote the mean packet size. Since
$1$~Mbps equals $1$~kbit/ms, the M/M/1 service rate of
link $e$ is $\mu_e=C_e/\bar{s}$ packets/ms.

An NSI $r=(s,t,B,L_{\max},T)$ specifies source/destination, required bandwidth, latency bound, and duration. $U_{\max}\in(0,1)$ is a network-wide link utilization threshold set by the operator. Suppose, on the National Science Foundation Network (NSFNet) backbone~\cite{matzner2021making}, an NSI requests a 100~Mbps flow from Salt Lake City to Atlanta with latency $\leq 52$ms and duration 30s, encoded as $r=(\text{Salt Lake City},\,\text{Atlanta},\,100\,\text{Mbps},\,52\,\text{ms},\,30\,\text{s})$.

Let $\mathcal{P}(s,t)$ denote the set of $K$ candidate paths between $s$ and $t$. A path $P\in\mathcal{P}(s,t)$ is \emph{feasible} iff:
\begingroup
\setlength{\abovedisplayskip}{3pt}
\setlength{\belowdisplayskip}{3pt}
\setlength{\jot}{1pt}
\begin{align}
  \forall\, e\in P:\ (1-\rho_e)\,C_e &\geq B \tag{C1}\\
  \forall\, e\in P:\ \rho_e + B/C_e &\leq U_{\max} \tag{C2}\\
  L(P)=\textstyle\sum_{e\in P}\ell_e &\leq L_{\max} \tag{C3}\\
  \forall\, k\in\mathcal{S}\ \text{s.t.}\ P_k\cap P\neq\emptyset:\ L'(P_k) &\leq L_k^{\max} \tag{C4}
\end{align}
\endgroup
where $\ell_e=d_e+
\rho_e^{\mathrm{new}}/[\mu_e(1-\rho_e^{\mathrm{new}})]$
is the sum of propagation delay and M/M/1 mean queueing delay, and $L'(P_k)$ is the projected latency of service $k$ after admission, and $\rho_{\max}(P)=\max_{e\in P}(\rho_e + B/C_e)$ is the maximum post-admission link utilization along $P$. The decision is to select $P^*=\arg\min_{P\in\mathcal{F}(r)}\sigma(P)$, where $\mathcal{F}(r)\subseteq\mathcal{P}(s,t)$ is the feasible set and
$\sigma(P)=\alpha \frac{L(P)}{L_{\max}} + (1-\alpha)\frac{\rho_{\max}(P)}{U_{\max}},\ \alpha\in[0,1]$
is a composite cost over normalized latency and utilization (lower is preferable), or reject with path-level reasoning if $\mathcal{F}(r)=\emptyset$.

\subsection{Underlying Deterministic \emph{what-if} Analysis for CAC}

\begin{figure}[t]
\centering
\resizebox{\columnwidth}{!}{%
\begin{tikzpicture}[
  proc/.style={
    rectangle, rounded corners=5pt,
    draw=blue!65!black, fill=blue!12,
    text width=2.6cm, align=center,
    minimum height=0.82cm,
    inner xsep=3pt, inner ysep=1.5pt,
    font=\small
  },
  datanode/.style={
    rectangle, rounded corners=3pt,
    draw=teal!65!black, fill=teal!10,
    text width=2.0cm, align=center,
    minimum height=0.68cm,
    inner xsep=3pt, inner ysep=1pt,
    font=\small\itshape
  },
  io/.style={
    rectangle, rounded corners=4pt,
    draw=gray!60!black, fill=gray!10,
    text width=5.8cm, align=center,
    minimum height=0.58cm,
    inner xsep=3pt, inner ysep=1.5pt,
    font=\small
  },
  decision/.style={
    rectangle, rounded corners=5pt,
    draw=orange!70!black, fill=orange!13,
    text width=5.8cm, align=center,
    minimum height=0.68cm,
    inner xsep=3pt, inner ysep=1.5pt,
    font=\small\bfseries
  },
  arr/.style={
    -{Stealth[length=4.5pt,width=3.5pt]},
    thick, blue!55!black
  },
  lbl/.style={
    font=\scriptsize\itshape,
    inner sep=1pt,
    text=gray!60!black
  }
]

%% Input
\node[io] (inp) at (0,0)
  {NSI $r=(s,t,B,L_{\max},T)$ + Network State $(G,\mathcal{S})$};

%% DT outer box
\begin{scope}[on background layer]
  \fill[blue!4, rounded corners=10pt]
    (-5.6,-0.65) rectangle (5.6,-5.95);
  \draw[
    blue!45!black,
    line width=1.2pt,
    rounded corners=10pt,
    dashed
  ]
    (-5.6,-0.65) rectangle (5.6,-5.95);
\end{scope}

%% Row 1: Pruner and pruned topology
\node[proc] (pruner) at (-3.45,-1.55)
  {\textbf{Bandwidth}\\[0pt]
   \textbf{Feasibility}\\[0pt]
   \textbf{Pruner (C1)}};

\node[datanode] (pruned) at (3.45,-1.55)
  {Pruned\\Network\\Topology};

\draw[arr] (pruner.east) --
  node[
    midway,
    below=2pt,
    font=\small\itshape,
    text=black
  ]
  {$\displaystyle
    (1-\rho_e)C_e < B
    \;\Rightarrow\;
    \text{prune}$}
  (pruned.west);

%% Row 2: K-shortest paths
\node[
  proc,
  fill=purple!10,
  draw=purple!60!black,
  text width=4.0cm
] (paths) at (0,-2.75)
  {\textbf{K-Shortest Candidate Paths}};

\draw[arr] (pruned.south)
  |- node[
    pos=0.35,
    right=4pt,
    font=\small\itshape,
    text=black
  ] {identify}
  (paths.north east);

%% Row 3: Three constraint checkers
\node[proc] (util) at (-3.8,-4.05)
  {\textbf{Link Util.}\\[0pt]
   \textbf{Constraint}\\[0pt]
   \textbf{Check (C2)}};

\node[proc] (lat) at (0,-4.05)
  {\textbf{Path Latency}\\[0pt]
   \textbf{Constraint}\\[0pt]
   \textbf{Check (C3)}};

\node[proc] (sla) at (3.8,-4.05)
  {\textbf{Affected Serv.}\\[0pt]
   \textbf{SLA}\\[0pt]
   \textbf{Check (C4)}};

\draw[arr] (paths.south)
  -- ++(0,-0.22) -| (util.north);
\draw[arr] (paths.south)
  -- (lat.north);
\draw[arr] (paths.south)
  -- ++(0,-0.22) -| (sla.north);

%% Row 4: Path scoring
\node[
  proc,
  fill=green!10,
  draw=green!60!black,
  text width=4.0cm
] (scoring) at (0,-5.30)
  {\textbf{Path Scoring \& Ranking}\\[0pt]
   $\sigma(P)$: latency,\ post-admission utilization};

\draw[arr] (util.south) |- (scoring.north west);
\draw[arr] (lat.south) -- (scoring.north);
\draw[arr] (sla.south) |- (scoring.north east);

%% Admission decision
\node[decision] (dec) at (0,-6.70)
  {Admission Decision:\\[0pt]
   \textsc{Admit}$(P^*)$ / \textsc{Reject}};

\draw[arr] (inp.south)
  -- ++(0,-0.18) -| (pruner.north);
\draw[arr] (scoring.south)
  -- (dec.north);

\end{tikzpicture}%
}
\caption{CAC pipeline with underlying \emph{what-if} analysis logic.
Six components (as shown in Table \ref{tab:models}) implement
Algorithm~\ref{alg:cac}, with constraint checks (C1--C4)}
\label{fig:pipeline}
\end{figure}
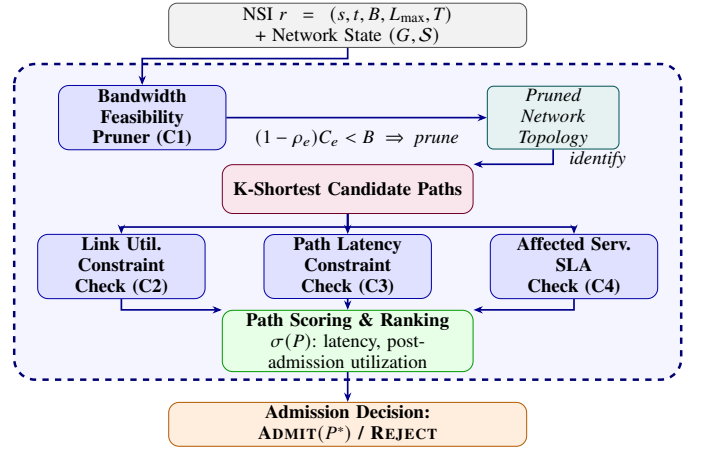

Fig.~\ref{fig:pipeline} illustrates the flow, and Algorithm~\ref{alg:cac} defines the deterministic \emph{what-if} analysis for CAC, comprising topology pruning (C1), path enumeration, constraint filtering (C2--C4), path selection, and state update. The formulation uses a rate-based utilization model and an M/M/1 queueing-delay model, where post-admission utilization and per-link delay are computed in Stages~1--3 of Algorithm~\ref{alg:cac} for constraints C1--C4. 
We have decomposed this logic into seven semantic models: six sub-components and one orchestrator (Table~\ref{tab:models}). In this case study, Go is used for its fast compilation, static typing, and self-contained binaries, although the system is language-agnostic.

\begin{algorithm}[H]
\caption{Deterministic \emph{what-if} analysis for CAC}
\label{alg:cac}
\begingroup
\scriptsize
\setlength{\algorithmicindent}{0.8em}
\renewcommand{\baselinestretch}{0.94}\selectfont
\begin{algorithmic}[1]
\REQUIRE Network state $(G,\mathcal{S})$; NSI $r=(s,t,B,L_{\max},T)$;\\
\hspace{1.8em}utilization threshold $U_{\max}$; number of candidate paths $K$
\ENSURE \textsc{Admit}$(P^*)$ or \textsc{Reject} with per-path reasons

\STATE \textit{Stage 1 -- Topology Pruning (C1):}
$G' \leftarrow \{e \in G : (1-\rho_e)C_e \ge B\}$

\STATE \textit{Stage 2 -- Path Enumeration:}
$\mathcal{P}(s,t) \leftarrow \mathrm{Yen}(G',s,t,K)$; $\mathcal{F}(r)\leftarrow\emptyset$

\STATE \textit{Stage 3 -- Constraint Filtering (C2--C4):}
\FORALL{$P \in \mathcal{P}(s,t)$}
    \IF{$\exists\, e \in P : \rho_e + \frac{B}{C_e} > U_{\max}$}
        \STATE Record reason: utilization violation (C2); \textbf{continue}
    \ENDIF

    \STATE $\rho_e^{\mathrm{new}} \leftarrow \rho_e + \frac{B}{C_e},\ \forall e\in P$
    \STATE $\ell_e \leftarrow d_e +
\frac{\rho_e^{\mathrm{new}}}
{\mu_e(1-\rho_e^{\mathrm{new}})}$,
for $\rho_e^{\mathrm{new}}<1$, else $+\infty$
    \STATE $L(P) \leftarrow \sum_{e\in P}\ell_e$

    \IF{$L(P) > L_{\max}$}
        \STATE Record reason: latency violation (C3); \textbf{continue}
    \ENDIF

    \STATE $\mathit{ok} \leftarrow \text{true}$
    \FORALL{$k \in \mathcal{S}$ such that $P_k \cap P \neq \emptyset$}
        \STATE Compute projected utilization $\rho_e^{\mathrm{proj}}$ and $L'(P_k)$
        \IF{$L'(P_k) > L_k^{\max}$}
            \STATE Record reason: SLA violation (C4); $\mathit{ok}\leftarrow\text{false}$; \textbf{break}
        \ENDIF
    \ENDFOR

    \IF{$\mathit{ok}$}
        \STATE $\mathcal{F}(r) \leftarrow \mathcal{F}(r) \cup \{P\}$
    \ENDIF
\ENDFOR

\IF{$\mathcal{F}(r) = \emptyset$}
    \STATE \textbf{return} \textsc{Reject}(reasons)
\ENDIF

\STATE \textit{Stage 4 -- Path Scoring:}
$P^* \leftarrow \arg\min_{P \in \mathcal{F}(r)} \sigma(P)$

\STATE \textit{Stage 5 -- State Update:}
\FORALL{$e \in P^*$}
    \STATE $\rho_e \leftarrow \rho_e + \frac{B}{C_e}$
\ENDFOR
\STATE Register $r$ in $\mathcal{S}$; \textbf{return} \textsc{Admit}$(P^*)$
\end{algorithmic}
\endgroup
\end{algorithm}

%

%\section{Discrete Event Simulation Framework}

\section{Experimental Evaluation}
\label{sec:evaluation}
Although architecturally LLM-backend-agnostic, the proposed system uses Claude Sonnet~4.6~\cite{anthropic_claude_2025} as the Coding-agent backend for the primary experiments and evaluates Qwen3-Coder-30B-A3B-Instruct in Section~\ref{qwen} to assess open-source-backend sensitivity.

\subsection{Sub-component Evaluation}

\subsubsection{Metrics}

Each generated Go program is evaluated using four metrics: compilation success $C \in \{0,1\}$, test pass rate $\mathcal{T} \in [0,1]$, code quality $Q \in [0,10]$, and semantic similarity $\xi \in [0,1]$. The overall correctness score is defined as
$S = w_1 C + w_2 \mathcal{T} + w_3 \tfrac{Q}{10} + w_4 \xi$,
where $w_1, w_2, w_3, w_4$ denote the relative importance of each metric with $\sum_i w_i = 1$. In our implementation, $w_1 = 0.30$, $w_2 = 0.50$, $w_3 = 0.15$, and $w_4 = 0.05$, prioritizing functional correctness over structural quality and semantic alignment. Code quality $Q$ is computed via Go Abstract Syntax Tree (AST) analysis with base 5.0 and increments for error handling (+2.0), function decomposition (+1.5), custom types (+1.0), and comments (+0.5). Semantic similarity $\xi$ is computed as a weighted comparison of JSON fields (30\%), data structures (20\%), operations (20\%), argument style (15\%), and computation logic (15\%).

\begin{table}[t]
\caption{CAC NDT: Semantic Models, Generated Code Size, Algorithmic Complexity, and Synthesis Cost (tokens, latency: mean $\pm$ std, 10 runs). $N{=}|V|$: nodes, $M{=}|E|$: links, $K$: candidate paths, $|P|$: path length, $|\mathcal{S}|$: active services, \textbf{LOC}: Lines Of Code}
\label{tab:models}
\centering
\scriptsize
\setlength{\tabcolsep}{2.5pt}
\renewcommand{\arraystretch}{0.95}

\rowcolors{2}{blue!8}{white}

\resizebox{\columnwidth}{!}{%
\begin{tabular}{p{1.6cm} c >{\raggedright\arraybackslash}p{3.1cm} p{1.7cm} c c}
\toprule
\rowcolor{blue!20}
\textbf{Semantic Model} & \textbf{LOC} & \textbf{Role in Algorithm~\ref{alg:cac}} & \textbf{Complexity} & \textbf{Tokens} & \textbf{Lat.\,(s)} \\
\midrule

Bandwidth Feasibility Pruner & 137 & C1: prune links, $(1-\rho_e)C_e < B$ & $\mathcal{O}(M)$ & $3059{\pm}79$ & $54.3{\pm}8.8$ \\

Graph Path Finder & 440 & $K$-shortest path algorithm (Yen's) & $\mathcal{O}(KN(M+N\log N))$ & $4758{\pm}145$ & $138.6{\pm}39.0$ \\

Link Utilization Constraint Checker & 169 & C2: $\rho_e^{\mathrm{new}} > U_{\max}$ & $\mathcal{O}(|P|)$ & $3517{\pm}383$ & $64.0{\pm}18.3$ \\

Path Latency Constraint Checker & 175 & C3: $L(P) > L_{\max}$ & $\mathcal{O}(|P|)$ & $3667{\pm}148$ & $52.2{\pm}21.3$ \\

Affected Services SLA Checker & 450 & C4: $L'(P_k) > L_k^{\max}$ & $\mathcal{O}(|\mathcal{S}|\cdot|P|)$ & $4071{\pm}113$ & $75.4{\pm}17.1$ \\

Dynamic Service Catalog & 465 & Path scoring $\sigma(P)$; state update & $\mathcal{O}(K\cdot|P|)$ & $4517{\pm}675$ & $107.7{\pm}47.7$ \\

\midrule
\rowcolor{green!15}
Orchestrator & 542 & Coordinates pipeline; final decision & $\mathcal{O}(KN(M+N\log N) + K|\mathcal{S}||P|)$ & $8813{\pm}173$ & $189.8{\pm}35.2$ \\
\bottomrule
\end{tabular}%
}
\end{table}

\subsubsection{Results}

\begin{table}[t]
\caption{Multi-run Code Generation Quality (mean $\pm$ std, 10 runs)}
\label{tab:components}
\centering
\scriptsize
\setlength{\tabcolsep}{2.5pt}
\renewcommand{\arraystretch}{0.95}

\rowcolors{2}{blue!8}{white}

\resizebox{\columnwidth}{!}{%
\begin{tabular}{lccccccc}
\toprule
\rowcolor{blue!20}
\textbf{Component} & \textbf{LOC} & \textbf{Attempts} & \textbf{Comp.} & \textbf{Test} & \textbf{Qual.} & \textbf{Sem.\ Sim.} & \textbf{Overall} \\
\midrule
Bandwidth Pruner & 137 & $2.1\pm0.3$ & 1 & 1.00 & $9.9\pm0.1$ & $0.76\pm0.02$ & $0.987\pm0.002$ \\
Path Finder & 440 & $2.3\pm0.4$ & 1 & 1.00 & $9.5\pm0.2$ & $0.77\pm0.02$ & $0.981\pm0.003$ \\
Util.\ Checker & 169 & $2.0\pm0.0$ & 1 & 1.00 & $10.0\pm0.0$ & $0.76\pm0.01$ & $0.988\pm0.001$ \\
Latency Checker & 175 & $2.0\pm0.0$ & 1 & 1.00 & $10.0\pm0.0$ & $0.76\pm0.01$ & $0.988\pm0.001$ \\
SLA Checker & 450 & $2.1\pm0.3$ & 1 & 1.00 & $10.0\pm0.0$ & $0.76\pm0.01$ & $0.988\pm0.001$ \\
Service Catalog & 465 & $2.1\pm0.3$ & 1 & 1.00 & $10.0\pm0.0$ & $0.76\pm0.01$ & $0.988\pm0.001$ \\
\midrule
\rowcolor{green!15}
\textbf{Average} & \textbf{306} & $2.1\pm0.2$ & \textbf{1} & \textbf{1.00} & $9.9\pm0.1$ & $0.76\pm0.01$ & $0.987\pm0.002$ \\
\bottomrule
\end{tabular}%
}
\end{table}

Table~\ref{tab:components} shows that all six sub-components compile and pass every semantic-model-defined test across 10 runs, establishing test-defined correctness. Attempts vary slightly due to differences in LLM convergence, while compilation and test outcomes remain invariant. Code quality and semantic similarity also remain stable, with the latter varying by only $\pm 0.01$--$0.02$, indicating consistent structure and semantic alignment. Similar scores are expected because several components derive structurally similar logic from uniform templates, with complexities from $O(|P|)$ to $O(|\mathcal{S}|\cdot|P|)$ (Table~\ref{tab:models}). The path-finding component exhibits slightly greater variance because Yen's $K$-shortest-path algorithm, with complexity $O(KN(M+N\log N))$, is structurally more demanding to synthesize.

Overall, the results show that the compile--test--debug loop produces reliable and repeatable implementations with Claude Sonnet~4.6. Table~\ref{tab:models} also reports mean token usage and wall-clock latency over the same 10 runs. From the Bandwidth Pruner ($O(M)$) to the Orchestrator ($O(KN(M+N\log N)+K|\mathcal{S}||P|)$), token usage increases from 3{,}059 to 8{,}813 (2.9$\times$), while latency increases from 54.3 to 189.8~s (3.5$\times$). Thus, synthesis cost increases moderately across the evaluated complexity range.

\subsubsection{LLM-backend sensitivity}\label{qwen}
We evaluated Qwen3-Coder-30B-A3B-Instruct~\cite{yang2025qwen3} on the six sub-component semantic models over 10 runs each. Claude Sonnet~4.6 succeeded in 60/60 runs, while Qwen produced compiling, test-passing implementations in 41/60 (68.3\%Catalogue): 10/10 for Dynamic Service Catalog, 9/10 each for Bandwidth Pruner, Path Latency Checker, and SLA Checker, 4/10 for Link Utilization Checker, and 0/10 for Graph Path Finder. The latter consistently generated only one of the two required distinct shortest paths, preventing orchestrator and complete CAC NDT synthesis. Across these models, per-model mean latency ranged from 86--488~s for Qwen versus 52--139~s for Claude Sonnet~4.6 (Table~\ref{tab:models}). Thus, the architecture supports different LLM backends, but synthesis reliability strongly depends on model capability.

\subsection{End-to-End CAC NDT Evaluation}
\label{sec:simulation}

\subsubsection{Simulation scenario}
A Python discrete-event simulator routes NSIs from arrival through the Intent handler and NDT factory to execution, scheduling arrivals and departures in a time-ordered priority queue. Correctness is evaluated against a handcrafted Python CAC NDT implementing Algorithm~\ref{alg:cac}, with both NDTs receiving the same arrival sequence and identical initial network and service states. We simulate 300 NSIs over 490.5~s on the 14-node, 21-link NSFNet~\cite{matzner2021making}, with Poisson arrivals (mean inter-arrival 1.71~s), $B\sim\mathcal{U}[50,200]$~Mbps, 33~s mean service duration, $L_{\max}\sim\mathcal{U}[30,100]$~ms, $U_{\max}=80\%$, and $K=3$. The weight $\alpha=0.6$ prioritizes latency as the service-critical dimension.

\begin{table}[t]
\caption{Behavioral Comparison: Synthesized vs. Reference CAC NDTs}
\label{tab:sim_small}
\centering
\scriptsize
\setlength{\tabcolsep}{2pt}
\renewcommand{\arraystretch}{0.95}

\rowcolors{2}{blue!8}{white}

\begin{tabularx}{\columnwidth}{
  >{\raggedright\arraybackslash}X
  >{\centering\arraybackslash}p{1.55cm}
  >{\centering\arraybackslash}p{1.55cm}
}
\toprule
\rowcolor{blue!20}
\textbf{Metric} & \textbf{Synth. NDT} & \textbf{Ref. NDT} \\
\midrule
Total requests
    & 300 & 300 \\
Admitted
    & 271 (90.3\%) & 273 (91.0\%) \\
Rejected
    & 29 (9.7\%) & 27 (9.0\%) \\
Blocking probability
    & 0.097 & 0.090 \\
\midrule
Bandwidth requested (Mbps)
    & 37{,}168.2 & 37{,}168.2 \\
Bandwidth admitted (Mbps)
    & 33{,}546.2 & 33{,}806.2 \\
Bandwidth efficiency
    & 90.3\% & 91.0\% \\
\midrule
\mbox{Avg. path latency of admitted NSIs (ms)}
    & 30.6 & 30.0 \\
\mbox{Avg. $L_{\max}$ of admitted NSIs (ms)}
    & 67.7 & 67.5 \\
\mbox{Avg. latency margin of admitted NSIs (ms)}
    & 37.0 & 37.4 \\
\midrule
\rowcolor{green!12}
Rejection cause
    & C3 only & C3 only \\
\rowcolor{green!12}
Decision agreement
    & 99.3\% & -- \\
\bottomrule
\end{tabularx}
\end{table}
\subsubsection{Results}
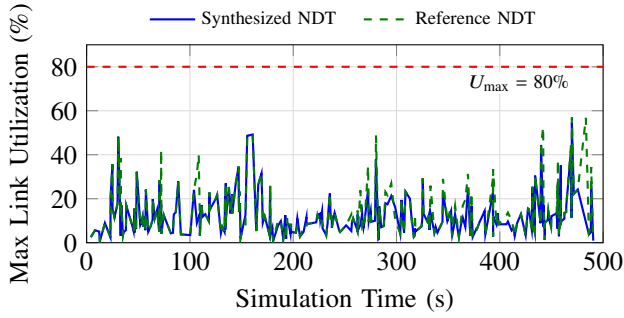
\begin{figure}[t]
\centering
\begin{tikzpicture}
\begin{axis}[
  xlabel={Simulation Time (s)},
  ylabel={Max Link Utilization (\%)},
  xmin=0, xmax=500,
  ymin=0, ymax=90,
  width=0.95\columnwidth,
  height=4.2cm,
  grid=both,
  grid style={line width=0.3pt, draw=gray!30},
  xtick={0,100,200,300,400,500},
  ytick={0,20,40,60,80},
  legend style={
    font=\scriptsize,
    at={(0.5,1.03)},
    anchor=south,
    legend columns=2,
    draw=none,
    fill=none,
    /tikz/every even column/.append style={column sep=8pt}
  },
  legend cell align={left},
]

\addlegendimage{blue, thick}
\addlegendentry{Synthesized NDT}
\addlegendimage{green!50!black, thick, dashed}
\addlegendentry{Reference NDT}
\addplot[blue, thick, mark=none] coordinates {
  (4.11,2.55) (8.11,5.69) (12.18,5.06) (12.81,0.71) (15.29,4.86)
  (17.71,8.99) (23.05,2.42) (23.18,20.47) (24.97,35.63) (25.09,15.58)
  (26.96,11.44) (29.92,17.78) (30.58,48.22) (32.68,11.82) (32.95,3.13)
  (33.10,12.48) (33.64,16.99) (35.18,3.41) (35.89,4.84) (38.02,5.72)
  (38.41,16.86) (44.77,6.77) (45.48,5.28) (46.26,15.51) (46.39,4.61)
  (46.70,12.78) (47.87,6.19) (48.53,32.29) (50.69,17.29) (51.90,7.97)
  (53.10,9.27) (54.20,11.71) (54.79,7.17) (55.34,11.97) (56.84,5.68)
  (57.35,24.13) (59.63,5.00) (62.01,7.53) (63.86,10.46) (63.98,19.85)
  (65.92,11.10) (68.24,15.05) (70.16,5.02) (70.64,28.40) (71.19,7.74)
  (71.47,9.95) (72.08,14.90) (72.11,17.50) (72.40,4.05) (74.65,12.57)
  (81.48,4.27) (83.61,4.50) (84.23,11.75) (84.45,12.92) (86.30,13.84)
  (88.65,27.88) (91.23,3.79) (100.31,3.48) (103.91,24.03) (106.73,8.99)
  (108.51,11.46) (109.52,13.03) (109.60,3.27) (111.24,17.14) (112.30,11.79)
  (114.88,12.98) (118.34,9.78) (118.91,22.80) (119.00,14.81) (124.95,19.93)
  (126.35,21.75) (129.99,4.32) (131.28,7.18) (133.53,4.86) (134.02,21.16)
  (134.25,5.96) (134.50,7.78) (134.70,27.02) (134.79,10.80) (137.73,25.32)
  (139.15,13.33) (140.75,22.10) (141.07,13.35) (144.35,26.99) (147.12,34.60)
  (147.47,13.94) (148.38,18.00) (149.03,2.11) (149.89,4.05) (153.74,12.52)
  (155.46,48.64) (160.73,49.20) (164.23,5.30) (166.15,26.46) (169.31,31.53)
  (170.55,9.05) (172.38,11.98) (173.03,10.92) (177.07,3.56) (177.31,11.87)
  (177.63,25.68) (178.16,4.05) (179.82,6.78) (180.53,0.85) (185.39,7.87)
  (188.03,9.84) (190.03,2.65) (190.32,1.58) (192.35,12.53) (193.13,1.90)
  (193.34,7.04) (194.03,11.09) (194.15,8.87) (195.71,8.31) (197.37,4.00)
  (197.88,8.59) (197.95,4.92) (202.62,4.57) (204.00,11.05) (206.55,2.82)
  (210.02,4.93) (212.82,11.60) (212.98,8.45) (222.27,9.27) (223.42,12.61)
  (225.81,13.20) (226.23,10.14) (227.91,14.41) (228.40,6.62) (231.95,3.39)
  (233.29,8.80) (234.19,15.33) (235.48,22.38) (236.29,11.27) (237.33,11.61)
  (237.38,6.82) (240.10,13.54) (244.39,5.68) (245.38,4.89) (251.20,7.89)
  (256.25,5.33) (256.88,9.23) (258.02,11.03) (262.19,5.63) (262.40,12.82)
  (262.73,5.77) (265.03,8.08) (266.71,13.22) (267.20,9.00) (267.96,6.61)
  (269.24,10.35) (272.49,24.30) (273.10,9.74) (275.01,14.54) (275.53,9.49)
  (279.18,10.01) (280.13,38.10) (282.42,1.52) (282.94,1.79) (283.26,16.56)
  (285.21,7.06) (287.64,7.58) (289.26,17.95) (291.29,17.47) (294.90,21.27)
  (298.76,15.99) (303.59,4.20) (304.72,16.81) (304.87,19.47) (307.28,4.57)
  (308.72,23.21) (312.90,20.10) (314.68,10.64) (314.92,10.21) (316.72,2.70)
  (317.69,3.32) (317.89,6.45) (318.36,5.11) (324.30,7.43) (324.99,11.30)
  (325.43,29.28) (326.19,13.20) (329.73,9.71) (331.59,13.05) (331.94,5.80)
  (332.95,13.24) (336.67,4.67) (337.84,7.29) (339.58,4.60) (344.48,9.48)
  (344.98,23.14) (349.25,9.80) (352.83,14.49) (353.39,5.44) (353.83,3.31)
  (355.61,7.49) (356.72,12.75) (360.40,2.62) (362.37,8.08) (364.16,11.40)
  (366.61,7.17) (369.18,24.71) (371.41,3.25) (372.16,13.89) (372.57,16.85)
  (373.78,2.63) (377.25,6.27) (377.80,6.14) (380.25,9.22) (385.33,14.47)
  (386.01,3.86) (393.37,22.79) (393.47,2.82) (393.54,12.00) (396.71,8.35)
  (397.82,13.41) (398.85,8.39) (406.44,8.39) (407.85,9.65) (411.45,5.56)
  (413.18,6.62) (414.45,3.07) (416.80,9.00) (420.46,12.97) (425.03,3.40)
  (427.30,15.53) (428.18,15.23) (431.11,5.50) (431.48,8.25) (431.58,25.43)
  (432.20,5.32) (432.32,8.21) (433.13,6.61) (433.96,4.66) (434.16,30.52)
  (439.43,8.86) (439.95,44.46) (442.00,7.37) (442.34,17.27) (443.21,6.96)
  (443.37,8.60) (443.57,1.23) (443.59,8.98) (446.35,10.36) (446.96,14.13)
  (448.14,9.93) (449.63,7.20) (449.93,12.08) (454.95,13.33) (455.88,6.41)
  (456.01,14.31) (456.20,26.96) (457.05,9.80) (457.26,2.94) (457.82,11.08)
  (459.09,35.22) (459.16,8.95) (459.78,9.89) (462.46,10.88) (462.59,13.55)
  (464.25,13.93) (464.33,16.67) (468.91,41.93) (469.35,11.22) (469.77,54.57)
  (471.63,22.17) (475.50,24.20) (486.26,3.78) (488.14,4.71) (488.47,26.03)
  (490.46,0.99)
};

\addplot[green!50!black, thick, dashed, mark=none] coordinates {
  (4.11,2.55) (8.11,5.69) (12.18,5.06) (12.81,0.71) (15.29,4.86)
  (17.71,8.99) (23.05,2.42) (23.18,20.47) (24.97,35.63) (25.09,15.58)
  (26.96,11.44) (29.92,17.78) (30.58,48.22) (32.68,11.82) (32.95,38.39)
  (33.10,12.48) (33.64,16.99) (35.18,3.41) (35.89,5.53) (38.02,5.72)
  (38.41,16.86) (44.77,6.77) (45.48,5.28) (46.26,15.51) (46.39,4.61)
  (46.70,12.78) (47.87,6.19) (48.53,32.29) (50.69,17.29) (51.90,7.97)
  (53.10,9.27) (54.20,11.71) (54.79,7.17) (55.34,11.97) (56.84,5.68)
  (57.35,24.13) (59.63,5.00) (62.01,7.53) (63.86,10.46) (63.98,19.85)
  (65.92,11.10) (68.24,15.05) (70.16,5.02) (70.64,28.40) (71.19,7.74)
  (71.47,9.95) (72.08,14.90) (72.11,42.19) (72.40,4.05) (74.65,12.57)
  (81.48,4.27) (83.61,4.50) (84.23,11.75) (84.45,12.92) (86.30,13.84)
  (88.65,27.88) (91.23,3.79) (100.31,3.48) (103.91,24.03) (106.73,31.36)
  (108.51,39.98) (109.52,13.03) (109.60,3.27) (111.24,17.14) (112.30,11.79)
  (114.88,12.98) (118.34,9.78) (118.91,22.80) (119.00,14.81) (124.95,19.93)
  (126.35,21.75) (129.99,4.32) (131.28,7.18) (133.53,4.86) (134.02,21.16)
  (134.25,5.96) (134.50,7.78) (134.70,27.02) (134.79,10.80) (137.73,25.32)
  (139.15,13.33) (140.75,22.10) (141.07,13.35) (144.35,26.99) (147.12,34.60)
  (147.47,13.94) (148.38,23.24) (149.03,2.11) (149.89,4.05) (153.74,7.92)
  (155.46,48.64) (160.73,49.20) (164.23,5.30) (166.15,26.46) (169.31,31.53)
  (170.55,9.05) (172.38,11.98) (173.03,10.92) (177.07,3.56) (177.31,11.87)
  (177.63,26.23) (178.16,4.05) (179.82,4.99) (180.53,0.85) (185.39,10.36)
  (188.03,9.84) (190.03,2.65) (190.32,1.58) (192.35,7.89) (193.13,1.90)
  (193.34,7.04) (194.03,11.09) (194.15,8.87) (195.71,8.31) (197.37,5.40)
  (197.88,8.59) (197.95,4.92) (202.62,4.57) (204.00,7.25) (206.55,2.82)
  (210.02,7.12) (212.82,6.95) (212.98,8.45) (222.27,9.27) (223.42,12.61)
  (225.81,8.56) (226.23,10.14) (227.91,9.76) (228.40,8.55) (231.95,3.39)
  (233.29,9.85) (234.19,15.33) (235.48,22.38) (236.29,11.27) (237.33,11.61)
  (237.38,6.82) (240.10,13.54) (244.39,5.68) (245.38,4.89) (251.20,7.89)
  (256.25,12.83) (256.88,9.23) (258.02,11.03) (262.19,5.04) (262.40,14.60)
  (262.73,5.77) (265.03,23.79) (266.71,7.47) (267.20,8.07) (267.96,6.61)
  (269.24,10.35) (272.49,34.88) (273.10,9.74) (275.01,14.54) (275.53,9.49)
  (279.18,10.01) (280.13,48.67) (282.42,1.52) (282.94,1.79) (283.26,16.56)
  (285.21,7.06) (287.64,7.58) (289.26,23.12) (291.29,21.65) (294.90,26.98)
  (298.76,10.16) (303.59,4.20) (304.72,19.86) (304.87,15.44) (307.28,4.57)
  (308.72,23.21) (312.90,20.50) (314.68,10.08) (314.92,10.21) (316.72,2.70)
  (317.69,3.32) (317.89,12.95) (318.36,5.11) (324.30,7.43) (324.99,14.93)
  (325.43,29.28) (326.19,13.20) (329.73,9.71) (331.59,13.05) (331.94,5.80)
  (332.95,25.93) (336.67,4.67) (337.84,6.18) (339.58,4.60) (344.48,9.38)
  (344.98,29.02) (349.25,9.80) (352.21,15.33) (352.83,14.49) (353.39,7.50)
  (353.83,3.31) (355.61,13.00) (356.72,16.79) (360.40,2.62) (362.37,10.14)
  (364.16,16.93) (366.61,19.51) (369.18,31.28) (371.41,3.25) (372.16,19.82)
  (372.57,22.78) (373.78,2.63) (377.25,6.27) (377.80,6.14) (380.25,9.22)
  (385.33,10.02) (386.01,3.00) (393.37,29.36) (393.47,2.82) (393.54,33.38)
  (396.71,8.35) (397.82,13.41) (398.85,8.39) (406.44,12.44) (407.85,13.70)
  (411.45,10.11) (413.18,6.62) (414.45,3.07) (416.80,9.00) (420.46,12.97)
  (425.03,3.40) (427.30,15.53) (428.18,15.23) (431.11,5.50) (431.48,8.25)
  (431.58,25.43) (432.20,5.32) (432.32,8.21) (433.13,6.61) (433.96,4.66)
  (434.16,30.52) (439.43,8.86) (439.95,44.46) (442.00,52.52) (442.34,17.27)
  (443.21,6.96) (443.37,8.60) (443.57,1.23) (443.59,8.98) (446.35,10.36)
  (446.96,14.13) (448.14,9.93) (449.63,7.20) (449.93,12.08) (454.95,13.33)
  (455.88,26.71) (456.01,14.31) (456.20,37.59) (457.05,9.80) (457.26,2.94)
  (457.82,11.08) (459.09,19.80) (459.16,8.95) (459.78,9.89) (462.46,10.88)
  (462.59,13.55) (464.25,13.93) (464.33,16.67) (468.91,44.29) (469.35,11.22)
  (469.77,56.93) (471.63,22.17) (475.50,24.20) (483.31,56.78) (486.26,4.57)
  (488.14,6.98) (488.47,34.43) (490.46,0.99)
};

\addplot[red, dashed, thick] coordinates {(0,80) (500,80)};

\node[
  font=\scriptsize,
  anchor=north east,
  yshift=-2pt,
  fill=white,
  inner sep=1pt
] at (axis cs:470,80) {$U_{\max}=80\%$};

\end{axis}
\end{tikzpicture}
\caption{Max link utilization (300 NSIs, 490.5~s) for both twins; peak 54.6\% (synth.) / 56.9\% (ref.), $U_{\max}\!=\!80\%$}
\label{fig:utilization}
\end{figure}

Table~\ref{tab:sim_small} summarizes the results for 300 NSIs. The synthesized Go NDT agrees with the handcrafted Python reference on 298 admission decisions (99.3\%), demonstrating close behavioral agreement. The two disagreements are latency-boundary cases: the path rejected in each case exceeds the corresponding maximum permitted latency by less than 1\%. Every admitted NSI satisfies constraints (C1)--(C4), while all rejections in both twins arise solely from C3. For state fidelity, comparing all 21 links after each of the 300 NSI decisions yields 6,300 utilization comparisons with a mean absolute error of 0.7 percentage points and a maximum of 24.4 points. This divergence arises mainly because numerical tie-breaking causes 22 of 271 jointly admitted requests to select different but equally admissible paths, leading subsequent loads to accumulate on different links. Fig.~\ref{fig:utilization} shows maximum link utilization over the 490.5,s simulation, peaking at 54.6\% for the synthesized NDT and 56.9\% for the reference, both below $U_{\max}=80\%$. Thus, C1, C2, and C4 are never binding, and the rise-and-fall pattern confirms correct resource allocation and release. However, this moderate-load scenario does not exercise the C2 boundary, motivating future higher-load validation. Because both twins process each NSI at the same event without artificial delay, state staleness is not evaluated.

%\paragraph{Network behavior}

\noindent\textbf{Security considerations:}
The Coding agent prompts the LLM using semantic models and Planner-assembled context, allowing malicious or modified semantic models or compromised context to inject instructions and produce unsafe code. Compilation and predefined tests detect syntax and tested behavioral errors but cannot guarantee freedom from vulnerabilities or untested behavior. Thus, ``verified'' means passing compilation and specified tests, not formal verification or security certification. Deployment additionally requires authenticated semantic models, isolated generation and testing, restricted tool access, static analysis, resource limits, and prior approval.

\section{Conclusion}
\label{conclusion}

In this paper, we presented the NDT factory, a multi-agent software system for on-demand synthesis of behavioral NDTs from semantic models. Validated on a CAC case study over NSFNet, the system generated a complete CAC NDT pipeline with 100\% compilation and test success across multiple runs. Simulation of 300 NSIs yielded 99.3\% decision agreement with a reference implementation and a 90\% admission rate, confirming reliable synthesis and deterministic execution.

Beyond CAC, the approach can extend to other behavioral network functions, such as quality of service (QoS) enforcement, fault localization, and traffic engineering, which can be expressed as semantic models and integrated into the same pipeline. Ongoing work includes systematic evaluation of LLM backends, semantic model refinement through execution feedback, dynamic intent classification, packet-level simulation, and extension toward 5G Radio Access Network (RAN) scenarios.
\section*{Acknowledgment}
This work is supported in part by MITACS Accelerate Program project IT43178 and in part by the Natural Sciences and Engineering Research Council (NSERC) DISCOVERY and CREATE TRAVERSAL programs.
%\balance

\bibliographystyle{IEEEtran}
\bibliography{references}

\end{document}